
\hbadness=10000
\vbadness=10000
\baselineskip=18pt
\magnification=1200
\line {May 1993 \hfil  INFN-CA-10-93}
 \vfill
\centerline {\bf DILATONIC BLACK HOLES IN THEORIES WITH MODULI FIELDS  }
\vskip 1.2in
\centerline {M. Cadoni}
\centerline {S. Mignemi}
\vskip .2in
\centerline {Istituto Nazionale di Fisica Nucleare, Sezione di Cagliari}

\centerline {Via Ada Negri 18, I-09127 Cagliari, Italy and Dipartimento}
\centerline {di Scienze Fisiche, Universit\`a di Cagliari}

\vfill
\centerline {\bf ABSTRACT}

We discuss the low-energy effective string theory when moduli of the
compactified manifold are present. Assuming a nontrivial coupling of the moduli
to the Maxwell tensor,  we find a class of regular black hole solutions.
Both the thermodynamical and the geometrical structure of these solutions
are discussed.

\eject
\def\d{{\rm d}}\def\ef{e^{-2\Phi}}\def\es{e^{-{2\over 3}
\Sigma}}\def\R{{\cal R}}\def\rie{{\cal R_{\mu\nu}}}\def\lag{{\cal L}}
\def\ez{e^{2\zeta}}\def\eza{\eta_0}\def\eua{\eta_1}\def\um{\left(1-
{r_-\over r}\right)}\def\up{\left(1-{r_+\over r}\right)}\def\ump{\left(
1-{r_-\over r_+}\right)}\def\a{{3\over 2q^2+3}}\def\b{{2q^2\over 2q^2+3}}
\def\c{{q^2\over 2q^2+3}}

{\noindent\bf 1. Introduction}

Electrically and magnetically charged black holes solutions arising in the
context of effective low-energy string theories exhibit properties which make
them drastically different from the usual Reissner-Nordstr\"om black holes
[1, 2]. When expressed in terms of the "string" metric,  these solutions are
characterised by a geometry which in the extremal limit is singularity free.
Furthermore the rather unusual thermodynamical properties seem to indicate
that black holes behave much more like elementary particles [3].
This situation is therefore very promising for trying to understand
long-standing puzzles of black holes physics such as the loss of quantum
coherence in black holes evaporation.

Most of the models considered until now work with the Einstein action modified
with the introduction of the dilaton field.
This is, from the point of view of the low-energy string action, just a first
approximation. For large mass holes $(M\gg M_{PL})$ this may be a good one.
However if string theory has to solve the puzzles of quantum gravity one needs
to go beyond this approximation. Various directions can be followed.
First, the low-energy string action has the form of an expansion in the inverse
string tension $\alpha'$. One should be able to take into account terms of the
action proportional to higher powers in the curvature tensors. Progress in this
direction has been made in Ref. [4].  Secondly,  even though one consider only
the massless sector of the string spectrum,  there are other fields, such
as the moduli of the compactified manifold, which should come into play.
Thirdly, correction to the effective low-energy action may appear at the
one-loop string level. Moreover nonperturbative effects should be considered.

The important question is: are the above-mentioned properties of dilatonic
black holes shared by more general situations?
Are they an artifact of the approximation or are they related to general
features of string theory?
In this paper we will move a step further in this direction. We
will derive and study black hole solutions of the low-energy string action
coming from dimensional reduction from ten to four dimensions, retaining
one single modulus which describes the radius of the compactified space.
By invoking one-loop string effects we will couple this field to the
Maxwell tensor in order to get new black hole solutions.

The structure of this paper is as follows. In section 2 we present the
low-energy effective string action we want to discuss. In section 3 we derive
the corresponding black hole solutions. A generalized version, together with
its black hole solutions, of our action is studied. Both the thermodynamical
and geometrical features of the solutions are discussed. In section 4 we
consider the duality symmetries of our theory and show how dual dyonic
solutions
can be generated.

{\noindent\bf 2. The action}

We start from the four-dimensional low-energy action of heterotic superstring
theory derived by Witten in Ref. [5]. The effective action was obtained through
dimensional reduction of ten-dimensional supergravity using a suitable
truncation of the spectrum. In the following we set to zero all the fields
but the graviton, the electromagnetic field, the dilaton $\phi$ and the
scalar $\sigma$ arising from the ansatz $ g_{IJ}=e^\sigma \delta_{IJ}$ for
the metric of the extra six dimensions. The action reads
$$A=\int\d^4x\sqrt{-g}\left[\R-6(\nabla\sigma)^2-8(\nabla\phi)^2
-e^{-2\phi+3\sigma}F^2\right].\eqno(1)$$
The corresponding field equations have regular black hole solutions only
for $\sigma=const.$ (these are the Garfinkle-Horowitz-Strominger
(GHS) solutions).
In fact performing the redefinitions
$$\Phi=\phi -{3\over2}\sigma,\qquad\qquad\Sigma=-{3\over2}\sigma-3\phi,
\eqno(2)$$
one can easily see that the action (1) becomes the one of Ref. [1] plus a
kinetic term for the field $\Sigma$. In particular there is no coupling of
$\Sigma$ to $F^2$. This prevents the appearance of new black hole solutions.
Note that $\Phi$ and $\Sigma$ rather than $\phi$ and $\sigma$ are the physical
dilaton and "compacton" fields of the four-dimensional supergravity theory.

At this stage one can ask himself if a coupling of the field $\Sigma$ to
$F^2$ could be justified. It is well-known that at the string tree level
such coupling does not exist. However it can be generated at the one-loop
level. A correction  to the action (1) of the form
$$\delta\lag \sim\es F^2\eqno(3)$$
was considered some time ago by Iba\~nez and Nilles, using arguments based
upon the supersymmetrization of anomaly cancelling terms [6].
More generally, terms of the type $\Delta(\es)F^2$ are expected to arise
from integrating out heavy modes of the string spectrum [7,8].
Furthermore, the appearance of such terms in the effective supergravity
lagrangian seems to be crucial in order to get dynamical supersymmetry
breaking [8].

For the moment we consider the "minimal" coupling defined by (3).
Later, we will study a more general situation. The form of the action we
are lead to study is then the following:
$$A=\int\d^4x\sqrt g\left[\R-2(\nabla\Phi)^2-{2\over 3}(\nabla
\Sigma)^2-\ef F^2-\es F^2\right].\eqno(4)$$

{\noindent\bf 3. The black hole solutions}

The field equations coming from the action (4) are:
$$\eqalign{\rie&=2\nabla_\mu\Phi\nabla_\nu\Phi+{2\over 3}\nabla_\mu
\Sigma\nabla_\nu\Sigma+2\left(\ef+\es\right)\left(F_{\mu\rho}F_{\nu\rho}-
{1\over 4}F^2g_{\mu\nu}\right),\cr
\nabla^2\Phi&=-{1\over 2}\ef F^2,\cr
\nabla^2\Sigma&=-{1\over 2}\es F^2.}\eqno (5)$$

A spherically symmetric solution of the field equations can be found by an
ansatz which reduces the system to a Toda-lattice form [2]:
$$\d s^2=e^{2\nu}(-\d t^2+e^{4\rho}\d\xi^2)+e^{2\rho}\d\Omega^2,$$
$$F_{mn}=Q\epsilon_{mn},\eqno(6)$$
where $\nu$ and $\rho$ are functions of $\xi$ and $Q$ is the magnetic charge.

Defining $\zeta=\nu+\rho$, the field equations are given by ($'=\d/\d\xi$):
$$\eqalign{\zeta ''=\ez,\cr
\phi ''=-Q^2e^{-2(\Phi-\nu)},\cr
\Sigma ''=-Q^2e^{-2(\Sigma/3-\nu)},\cr
\nu ''+\Phi ''+\Sigma ''=0,}\eqno(7)$$
with the constraint
$$\zeta '^2-\nu '^2-\Phi '^2-{1\over 3}\Sigma '^2+Q^2e^{-2(\Phi-\nu)}+Q^2
e^{-2(\Sigma/3-\nu)}-\ez=0.\eqno (8)$$

An exact solution can be found if $\Sigma '=3\Phi '$. By introducing a
new variable, such that $\ez=\eta^2-\eza^2$, the only asymptotically flat
solution with a regular horizon can be written as:
$$e^{2\nu}=(\eta -\eza)^{3/5}(\eta+\eza)(\eta+\eua)^{-8/5},$$
$$e^{2\rho}=(\eta -\eza)^{2/5}(\eta+\eua)^{8/5},\eqno(9)$$
$$e^{-2\Phi}=(\eta-\eza)^{2/5}(\eta+\eua)^{-2/5},$$
where $\eza$ and $\eua$ are integration constants.

The solution assumes a neater form by defining a new variable $r$ such that
$r=\eta +\eua$, $r-r_+=\eta+\eza$, $r-r_-=\eta-\eza$. As we shall see, the
constants $r_+$
and $r_-$ are simply related to the physical mass $M$ and charge $Q$ of the
black hole. One has now:
$$\d s^2=-\lambda^2 \d t^2+\lambda^{-2}\d r^2+R^2\d\Omega^2,\eqno(10)$$
with
$$\lambda^2=\up\um^{3\over 5},\qquad\qquad R^2=r^2\um^{2\over 5},\eqno(11.a)$$
and
$$e^{-2\Phi}=\um^{2\over 5}.\eqno(11.b)$$

The action (4) corresponds to the "minimal" coupling (3). However, one can
also consider more general couplings:
$$\delta\lag=e^{-{2q\over 3}\Sigma}, \qquad q\in R.\eqno(12)$$
These terms may be viewed as the building blocks of a series expansion of
the general term $\Delta(\es)F^2$. The action is now:
$$A=\int\d^4x\sqrt{-g}\left[\R-2(\nabla\Phi)^2-{2\over 3}(\nabla
\Sigma)^2-\ef F^2-e^{-{2q\over 3}\Sigma} F^2\right].\eqno(13)$$
Proceeding as before, one readily obtains a regular black hole solution
for  $\Sigma '=3q^{-1}\Phi '$. The solutions are:
$$\lambda^2=\up\um^\a,\eqno(14.a)$$
$$R^2=r^2\um^\b,\eqno(14.b)$$
$$\ef=\um^\b.\eqno(14.c)$$
The metric functions of our solution coincide, after the redefinition $a^2
=q^2(q^2+3)^{-1}$, with that found in Ref. [1] in the context of a
generalized model for dilaton gravity. The expression (14.c) for the dilaton,
however, is not the same.
As we shall see later, this has important consequences for the form of the
"string" metric.

The physical mass and charge of the hole are given by:
$$2M=r_++\a\, r_-,\qquad\qquad Q^2=\c\, r_+r_-.\eqno(15)$$
Using well-known formulae one can easily calculate the temperature and the
entropy of the black hole. We have:
$$T={1\over 4\pi r_+}\ump^\a,\eqno(16)$$
$$S=\pi r_+^2\ump^\b.\eqno(17)$$
We see that for extremal black holes $(r_+\rightarrow r_-)$ both $T$ and
$S$ approach monotonically to zero. This behaviour is different from that
of GHS-dilatonic black holes for which the temperature approaches to a constant
value in the extremal limit. Thus the interpretation of the final state of
the black holes described by (14) is straightforward: zero entropy at zero
temperature indicate a non degenerate ground state,  which naturally does not
radiate.
In the extremal limit the spacetime still displays a naked singularity.
However the metric appearing in the string sigma-model is not $g_{\mu\nu}$
but rather $e^{2\Phi}g_{\mu\nu}$. This is the metric to which strings couple.
The charged extremal black hole metric (10) becomes now
$$ds^2_{string}=-\up^{6/(2q^2+3)}dt^2+\up^{-2}dr^2+r^2d\Omega^2.\eqno(18)$$
The geometry of the $t=const$ surfaces is identical to that of the extremal
GHS-dilatonic black hole. There is a semi-infinite throat attached to an
asymptotic
flat region. The only difference resides in the form of the $g_{tt}$ component
of the metric. Whereas in the GHS case $g_{tt}=const$ in eq. (18) $g_{tt}
\rightarrow 0$ as $r\rightarrow r_+$. This indicates that the horizon though
infinitely far away along a space-like geodesic is not such for a time-like
one.

{\noindent\bf 4. Dual solutions}

It is well known that in the case of GHS-dilatonic black holes one can exploit
the invariance of the field equation under $SL(2,R)$ to generate dual dilaton
dyons [9,10]. This invariance holds, though in a restricted sense, also for the
equation of motion coming from an improved form of the action (4).
To show this let us introduce in the action (4) the two axion fields $a$
and $D$ coming from the ten-dimensional axion $B_{IJ}$, together with the
couplings of these fields to $F\tilde F$ (see Ref. [5]).
Defining the complex scalar fields
$$\eqalign{\rm T=3\sqrt 2 D+ i\ef,\cr {\rm S}=\sqrt 2 a + i\es,\cr}\eqno(19)$$
the action can be written as
$$A=\int\d^4x\sqrt{-g}\left\{\R- G_{i\bar k}\partial_\mu Z^i\partial_\nu {\bar
Z}^{\bar k}
g^{\mu\nu} +{i\over 4}\left[\left({\rm S}+{\rm T}\right) F_+^2-\left(\bar
{\rm S}+\bar {\rm T}\right) F_- ^2 \right]\right\},\eqno(20)$$
where $G_{i\bar k}$ is the K\"ahler metric of $SL(2,R)/U(1)\times SL(2,R)/U(1)$
corresponding to the K\"ahler potential $G=-\ln i\left({\rm S}-\bar {\rm S}
\right)-
3 \ln i\left({\rm T}-\bar {\rm T}\right)$, $Z_i\doteq\left\{{\rm S},{\rm T}
\right\},
i=1,2$
and $F_\pm =F\pm i \tilde F$.

The sigma-model kinetic terms of the action (20) are invariant under the
non-linearly realised $SL(2,R)\times SL(2,R)$ group. The last two terms of the
action
however break this symmetry. Nevertheless,  one can easily verify, by writing
down the field equations,  that along the orbit of solutions verifying
${\rm T}=3{\rm S}$,  the field equations are invariant under the following
$SL(2,R)$
transformations
$${\rm S}\rightarrow {a {\rm S}+b\over c {\rm S}+d},\qquad F_+\rightarrow -
\left(c {\rm S}+ d\right)F_+,\qquad ad-bc=1.\eqno(2')$$
Using the  invariance of the field equations under the former transformations,
one can generate from the solutions (11) dual dyon solutions. These are
similar to the ones obtained in Ref. [10]. One needs just to use there the
expression (11b) for the dilaton and $\Phi'=3\Sigma'$ to get the corresponding
expression for $\Sigma$.

{\noindent\bf Conclusions}

In this paper we have found and analysed black hole solutions of dilaton
gravity
when a non trivially-coupled modulus of the compactified manifold is present.
Our solutions share common properties with the GHS-dilatonic black holes.
In particular the spatial geometry associated with the "string" metric in the
extremal limit is the same. The full spacetime geometry and the thermodynamical
properties are,  however,  different. A deeper insight into the whole subject
could be achieved by studying the corresponding two-dimensional gravity theory.
We plan to discuss this topic in a forthcoming paper.

\smallskip
\centerline {\bf References}
\smallskip

[1] D. Garfinkle, G.T. Horowitz and A. Strominger, Phys. Rev. {\bf D43} (1991)
3140.

\hangindent=0,45in \hangafter=1
[2]  G.W. Gibbons and K. Maeda, Nucl. Phys.
     {\bf B298} (1988) 748.

\hangindent=0,45in \hangafter=1
[3] C.F.E. Holzhey and F. Wilczek, Nucl. Phys. {\bf B380} (1992) 447.

\hangindent=0,4in \hangafter=1
[4] S. Mignemi and N.R. Stewart, Phys. Rev. D (in press).

\hangindent=0,4in \hangafter=1
[5] E. Witten, Phys. Lett. {\bf B155} (1985) 151.

\hangindent=0,4in \hangafter=1
[6] L. E. Iba\~nez and H.P. Nilles, Phys. Lett. {\bf B169}
(1986) 354.

\hangindent=0,4in \hangafter=1
[7]  V. Kaplunovsky,  Nucl. Phys. {\bf B307} (1988) 145.

\hangindent=0,4in \hangafter=1
[8] J. Dixon, V. Kaplunovsky, J. Louis, Nucl. Phys. {\bf B355} (1991) 649.

\hangindent=0,4in \hangafter=1
[9] J. Schwarz, Caltech Preprints CALT-68-1815; A. Sen, Preprints
TIFR-TH-92-57.

\hangindent=0,4in \hangafter=1
[10] A. Shapere, S. Trivedi and F. Wilczek, Mod. Phys. Lett. {\bf A29} (1991)
2677.

\end